\documentstyle[12pt]{article}

\def\la{\langle}
\def\ra{\rangle}

\def\lb{\lbrack}
\def\rb{\rbrack}

\def\wt{\widetilde}

 \def\Slash#1{
  \begin{picture}(5,6)(0,0)
  \put(-.7,-1.2){\line(5,6)6}
  \end{picture}
  \kern-.8em#1}
 \def\slash#1{
  \begin{picture}(5,6)(0,0)
  \put(-1.5,-1.7){\line(5,6)5}
  \end{picture}
  \kern-.8em#1}

\def\Sn{\Slash \nabla}
\def\sd{\Slash \partial}

\def\Tr{\mbox{Tr}}
\def\tr{\mbox{tr}}
\def\e{\epsilon}
\def\g5{\gamma_5}
\def\hg5{\hat{\gamma}_5}
\def\A{{\cal A}}
\def\index{\mbox{index}}

\def\R{{\cal R}}
\def\O{{\cal O}}

\def\seu{\sum_{\mu}(1-\epsilon_{\mu}\sqrt{1-s_{\mu}^2}\,)}

\def\Qlatmr1{Q_{lat}^{(m=r=1)}}

\def\be{\begin{eqnarray}}
\def\ee{\end{eqnarray}}
\def\td{\widetilde}
\def\ttd{\tilde}

\makeatletter
\def\section{\@startsection{section}{1}{\z@}{-3.5ex plus -1ex minus -.2ex}
{2.3ex plus .2ex}{\large\bf}}
\makeatother

\makeatletter
\@addtoreset{equation}{section}

\makeatother

\parskip=\medskipamount

\begin{document}



\begin{center}

{\large \bf Axial anomaly and topological charge in lattice gauge theory
with overlap Dirac operator}
\\

\vspace{1ex}

{\large David H. Adams}

\vspace{0.5ex}

Math. dept. and Centre for the Subatomic Structure of Matter, \\
University of Adelaide, S.A. 5005, Australia. \\


email: dadams@maths.adelaide.edu.au

\end{center}

\begin{abstract}

An explicit, detailed evaluation of the classical continuum limit of the
axial anomaly/index density of the overlap Dirac operator is carried out
in the infinite volume setting, and in a certain finite volume setting where
the continuum limit involves an infinite volume limit.
Our approach is based on a novel power series expansion of
the overlap Dirac operator.
The correct continuum expression is reproduced when the parameter $m_0$
is in the physical region $0<m_0<2$. This is established for a broad 
range of continuum gauge fields.
An analogous result for the fermionic topological charge, given by the index 
of the overlap Dirac operator, is then established
for a class of topologically non-trivial fields in the aforementioned
finite volume setting.
Problematic issues concerning the index in the infinite volume setting are
also discussed.

\end{abstract}

\section{Introduction}

In the last few years there have been interesting developments 
in lattice gauge theory
concerning chirality aspects of lattice fermions and topological aspects
of lattice gauge fields. The overlap formulation of chiral gauge theories
\cite{ov1,ov2}, which was largely inspired by Kaplan's domain wall 
proposal \cite{Kaplan}, led to the introduction of the overlap Dirac
operator for massless lattice fermions \cite{Neu(PLB)}. This operator
satisfies the the Ginsparg--Wilson relation \cite{GW}, thereby providing 
an explicit solution to the chirality problem in lattice QCD \cite{L(PLB)}
(see, e.g., \cite{review} for reviews).

A key test of any lattice formulation of a quantum field theory is whether it
reproduces the correct anomalies for the continuum symmetries.
In particular, the axial anomaly in QCD explains the $\pi^0\to2\gamma$
decay amplitude \cite{axial} and should be reproduced in the lattice
formulation. Another important quantity in lattice gauge theory is the
topological charge of lattice gauge fields. The topological charge enters
into an explanation for baryon number violation in the standard model,
and for the relatively large value of the $\eta'$
mass in QCD \cite{baryon-eta}. Lattice definitions of topological charge is
therefore an interesting topic of study. A necessary condition for an 
acceptable lattice definition of topological charge is that it has the 
correct continuum limit.

In this paper we study the classical continuum limit of the axial anomaly
and fermionic topological charge in lattice gauge theory with fermion
action specified by the overlap Dirac operator. 
Our approach avoids the limitations of earlier approaches which were only
valid for small (and hence topologically trivial) background gauge fields.
In \S3 we develop new techniques (first introduced in \cite[(v4)]{DA} and
\cite{Taiwan}) for evaluating the classical continuum limit of the 
index density/axial anomaly, which enable us to rigorously
show that the correct continuum expression is obtained (when $0<m_0<2$) 
for a broad range of continuum gauge fields. These include topologically
non-trivial fields and more general classes of fields which may diverge
at infinity or have singularities.

The situation regarding the index 
is more delicate. In the continuum, in infinite volume,
the index of the Dirac operator coupled to an $SU(n)$ gauge field 
$A(x)=A_{\mu}^a(x)T^adx^{\mu}$ on ${\bf R}^4$ is ill-defined in general.
A related fact is that the topological charge,
\be
Q(A)=\int_{{\bf R}^4}d^4x\,q^A(x)\quad,\quad\ 
q^A(x)=\frac{-1}{32\pi^2}\e_{\mu\nu\rho\sigma}\tr(
F_{\mu\nu}(x)F_{\rho\sigma}(x))\,,
\label{1}
\ee
is not integer valued,\footnote{
Examples of fields with arbitrary non-integer topological charge are
easily constructed: View ${\bf R}^4$ as ${\bf R}^2\times{\bf R}^2\,$,
let $(r_1,\theta_1)$ and $(r_2,\theta_2)$ be polar coordinates on
the two copies of ${\bf R}^2$ ($\theta_j\in[0,2\pi]$) and choose a smooth
function $\lambda(r)$ on ${\bf R}_+$ which vanishes in a neighbourhood of
$0$ and is equal to a constant $c$ for $r\ge{}r_0$ (for some finite $r_0$).
Then for any generator $T$ of $SU(N)$ the field 
$A=\lambda(r_1)Td\theta_1\pm\lambda(r_2)Td\theta_2$ has topological charge
$Q(A)=\mp\tr(T^2)c^2$. This field and its partial derivatives are bounded 
on ${\bf R}^4$. Note that the non-integrality of $Q(A)$ does not contradict
the results of \cite{Uhlenbeck} since the Euclidean action 
of the field is infinite.}
or even well-defined,\footnote{Even
if one allows the values $\pm\infty\,$, $Q(A)$ is still not well-defined 
in general since, e.g., the integrand in (\ref{1}) 
can be oscillatory at infinity.} 
in general. 
However, integer values for $Q(A)$ are guaranteed when $A$ converges
sufficiently quickly to a pure gauge configuration at infinity (or at the
singularity in the case where $A$ is in a singular gauge). It is in this 
case that the Dirac operator coupled to $A$ has a well-defined index equal
to $Q(A)$ by the Index Theorem (since in this case $A_{\mu}(x)$
corresponds to a gauge potential in an $SU(n)$ bundle over the compact 
manifold $S^4\cong{\bf R}^4\cup\{\infty\}$ \cite{Uhlenbeck}).
Similarly, in the lattice theory the index of the overlap Dirac operator
$D^U$ will be ill-defined in general in the
infinite volume setting (i.e. infinite lattice on the whole of ${\bf R}^4$).
One expects that, in analogy with the continuum case, 
$\index\,D^U$ will be a well-defined finite number when suitable
conditions are imposed on the lattice gauge field $U$, e.g. a requirement
that it converge sufficiently quickly to a pure gauge configuration
at infinity. A first step towards clarifying this issue is carried out 
in \S4, where other subtleties/pitfalls concerning $\index\,D^U$ 
in the infinite volume setting are also discussed.\footnote{In \cite{K} it 
was erroneously claimed that $\index\,D^U\to{}Q(A)$ in the classical
continuum limit if only $A_{\mu}(x)$ and the derivatives 
$\partial_{\nu}A_{\mu}(x)$ are bounded.
We emphasize that this is not true in general: these conditions are
not enough to guarantee that $Q(A)$ is integer or even well-defined, 
or that $\index\,D^U$ is well-defined, cf. the example above.}

To confirm that $\index\,D^U$ really is able to capture topological
information about the continuum gauge field
it is desirable to verify that $\index\,D^U$ reduces to $Q(A)$
in some setting where the former is a well-defined integer from the beginning
and the problems alluded to above do not arise.
To do this, in \S5 we consider a setting where the continuum
field $A_{\mu}(x)$ is in a singular gauge
and vanishes outside a bounded region of ${\bf R}^4$. We restrict the lattice
transcript $U$ of $A$ to a large finite volume hypercube; then, with 
(anti)periodic boundary conditions on the lattice spinor fields,
$\index\,D^U$ is a well-defined finite integer. We show that, in the
continuum limit defined by first taking
the infinite volume limit and then the $a\to0$ limit, $\index\,D^U$ reduces 
to $Q(A)$.
Since $\index\,D^U$ and $Q(A)$ are both integers in this setting it follows 
that they are also equal close to this continuum limit, i.e. for all 
sufficiently large volumes and small lattice spacings. This analytic result
complements the previous numerical results of \cite{Edwards}.

The classical continuum limits of the axial anomaly and topological charge
in the finite volume 4-torus case are not considered here; this case is 
treated in a separate paper \cite{DA(torus)}.

The organisation of this paper is as follows. \S2 provides the relevant
background material. The material in \S3--5 has been described above.
(\S3: classical continuum limit of the index density/axial anomaly;
\S4: aspects of $\index\,D^U$ and its continuum limit in the infinite
volume lattice setting; \S5: classical continuum limit of $\index\,D^U$
starting in a finite volume lattice setting and taking an infinite volume 
limit.) The conclusions of the paper are summarised in \S6.

\section{Background}

In continuum QCD the axial anomaly arises from the triangle diagram
in perturbation theory \cite{axial}, and can also be understood 
non-perturbatively as arising from the jacobian for chiral transformations
of the fermion fields, regularised by Fujikawa's method \cite{Fuji(cont)}.
In traditional lattice fermion formulations, such as Wilson-Dirac
and staggered fermions, the axial anomaly arises in a different way:
The chiral jacobian is trivial, and the anomaly arises instead as a result of
the breaking of chiral symmetry by the lattice fermion action; see, e.g.,
\cite{Smit,Seiler}. (The lattice fermion action needs to break chiral
symmetry in order to avoid species doubling \cite{NN}.)

However, a new perspective is possible for the new lattice fermion actions 
where the lattice Dirac operator satisfies\footnote{
This and the following structure is implicit in the vector overlap 
formulation; see, e.g., \cite{Kiku(overlap)}.}
\be
D\g5+\g5D&=&aD\g5D\qquad\qquad\ \mbox{(Ginsparg--Wilson relation \cite{GW})}
\label{2.1} \\
D^*&=&\g5D\g5\qquad\qquad\qquad\mbox{($\g5$--hermiticity)}
\label{2.1a}
\ee
($a$=lattice spacing). Besides the overlap Dirac operator 
((\ref{2.9}) below), another solution is the Dirac operator obtained in
the perfect action approach, which is given less explicitly via recursion 
relations
\cite{perfect}. The fermion action 
$S=a^4\sum_x\bar{\psi}(x)D\psi(x)$ is invariant under a new lattice-deformed
chiral transformation of the fermion fields \cite{L(PLB)},\footnote{More 
generally the action is invariant under
$\delta\psi=\g5(1-(1-t)aD)\psi\,$, $\delta\bar{\psi}=\bar{\psi}\g5(1-taD)\,$
$\forall\,t\in[0,1]$. These transformations all lead to the same axial
anomaly (\ref{2.4}).}
\be
\delta\psi=\hg5\psi\qquad\ ,\qquad\ \delta\bar{\psi}=\bar{\psi}\g5
\label{2.2}
\ee
where
\be
\hg5=\g5(1-aD)\,,
\label{2.3}
\ee
It follows from (\ref{2.1})--(\ref{2.1a}) that
$\hg5^2=1$ and $\hg5^*=\hg5\,$,
so $\hg5$ determines an orthogonal decomposition
of lattice spinor fields into ``chiral'' subspaces just like $\g5$.
The fermion measure is not invariant under the new chiral transformation
(\ref{2.2}), and the axial anomaly can be determined from the corresponding
jacobian to be \cite{L(PLB)}\footnote{This is the same as the axial
anomaly for the vector overlap, which was considered in a special case
in \S10 of \cite{ov2}.}
\be
\A(x)=\tr(\hg5(x,x))=-ia\tr(\g5D(x,x))
\label{2.4}
\ee
Thus the lattice regularisation with the new lattice fermion actions
is providing a lattice version of Fujikawa's non-perturbative perspective
on the axial anomaly.

A key question now is whether the axial anomaly (\ref{2.4}) for the new 
lattice fermion actions reduces to the usual axial anomaly
$\A_{cont}(x)=2iq^A(x)$ in the classical continuum limit. 
There have been a number of attempts to show this in the perturbative
framework for general lattice Dirac operator satisfying eq.'s 
(\ref{2.1})--(\ref{2.1a}) \cite{GW,laliena,Chiu-H}. A very general
perturbative argument which encompasses these lattice Dirac operators
has been given in \cite{Rothe}, with the correct continuum anomaly being 
reproduced provided certain general conditions are satisfied.
In the specific case of the overlap Dirac
operator an explicit perturbative calculation was carried out in
\cite{KY}. Subsequently,
other calculations of the continuum limit of the axial anomaly for the 
overlap Dirac operator were carried out in 
\cite[v1]{DA},\cite{Fuji},\cite{Suz}. These did not involve an explicit
perturbative expansion in the gauge field; however they did rely on the
gauge field being sufficiently small (and hence topologically trivial)
so that power series expansions of 
certain operators involving the gauge field could be carried out.
The arguments also required the gauge field to have a well-defined Fourier
transformation, which is not always true for general gauge field 
on ${\bf R}^4$.
Thus the problem of providing a non-perturbative derivation, valid for general
classes of gauge fields (including topologically non-trivial fields)
still remained. To provide such a derivation is one of the main
purposes of the present paper. Our explicit, detailed derivation of the 
continuum limit of the axial anomaly (in \S3)
complements an abstract, implicit argument in the case of
general gauge field outlined by M. L\"uscher in \S5 of 
\cite{L(nonabelian)}. (We should also mention that
the continuum limit of the axial anomaly in the vector version of the
overlap formulation was calculated in \cite{Strathdee}, and in the domain
wall formulation in \cite{Shamir}, although these calculations were based
on certain approximations (e.g. linearisation of the overlap) and/or
assumptions (e.g. weak field, slowly varying field).)

Another appealing feature of lattice Dirac operators satisfying the GW
relation (\ref{2.1}) is that the zero-modes have definite chirality,
i.e. the nullspace $\ker{}D$ is invariant under $\g5\,$: if
$D\psi=0$ then $D(\g5\psi)=(-\g5{}D+aD\g5{}D)\psi=0$ \cite{laliena}.
Thus, at least in finite volume settings, 
$\index\,D=\Tr(\g5|_{\ker{}D})$ is well-defined.
This is in contrast to traditional operators like the Wilson--Dirac
operator for which only an approximate (non-integer) index can be
defined \cite{Smit}. Furthermore, there is a ``lattice
index theorem'' \cite{laliena,L(PLB)}:
\be
\index\,D=-\frac{a}{2}\Tr(\g5{}D)\,.
\label{2.6}
\ee
Thus $\index\,D=a^4\sum_xq(x)$ with the index density
\be
q(x)=-\frac{a}{2}\tr(\g5{}D(x,x))
\label{2.7}
\ee
related to the axial anomaly (\ref{2.4}) by
\be
\A(x)=2iq(x)
\label{2.8}
\ee
just as in the continuum.
The index of $D$ provides a fermionic definition of the topological
charge of the background lattice gauge field. In the case of the 
overlap Dirac operator\footnote{In fact all solutions of 
(\ref{2.1})--(\ref{2.1a}) have this ``overlap'' form: $D$ is a solution
if and only if $D=\frac{1}{a}(1+\g5\e)$ for hermitian $\e$ ($=-\hg5$)
with $\e^2=1\,$,
cf. Neuberger's first paper in \cite{review}.}
\be
D=\frac{1}{a}\Bigl(\,1+\g5\e(H)\Bigr)\qquad,\quad\ \e(H)=
\frac{H}{\sqrt{H^2}}=-\hg5
\label{2.9}
\ee
where
\be
\frac{1}{a}H=\g5(D_w-r\frac{m_0}{a})
\label{2.10}
\ee
with $D_w$ being the Wilson--Dirac operator and $r>0$ the Wilson parameter,
the index formula (\ref{2.6}) gives\footnote{This is assuming a finite 
lattice and/or suitable periodicity conditions so that the space of lattice
spinor fields is finite-dimensional and hence $\Tr(\g5)=0$.
The role of the relation $\Tr(\g5)=0$ in this setting has been discussed
in detail in \cite{Fuji(PRD)}; related issues are discussed in 
\cite{Chiu(PRD1+2),Yamada}.}
\be
\index\,D=-\frac{1}{2}\Tr\Bigl(\,\frac{H}{\sqrt{H^2}}\Bigr)
\label{2.11}
\ee
i.e. the spectral asymmetry of $-H=-H_{m_0}$. 
Since $H_m$ has symmetric spectrum
and no zero-modes for $m<0$ \cite{ov1,ov2,Edwards}, (\ref{2.11}) equals
minus the spectral flow of $H_m$ coming from the eigenvalue crossings that
occur in the interval $0\le{}m<m_0$. This is precisely the definition of
lattice topological charge arising in the overlap formulation \cite{ov1,ov2}.
(The spectral flow definition of lattice topological charge had earlier
been studied numerically in \cite{Itoh}.) This lattice topological charge
is well-defined for all lattice gauge fields except those for which
$H_{m_0}$ has zero-modes. After excising this measure-zero subset, the
space of lattice gauge fields splits into topological sectors labelled by
$\index\,D$. (This is reminiscent of the way that L\"uscher's geometrical
definition of lattice topological charge gives a decomposition into 
topological sectors after excising a measure-zero subset \cite{L(CMP)}.)
The physical region for the parameter $m_0$, i.e. the values for which
$D$ is not aflicted with species doubling in the fee fermion case, 
is $0<m_0<2$ \cite{ov1,ov2,Neu(PLB)}. 
As we will see, it is precisely for $m_0$ in
this region that the axial anomaly has the correct classical continuum
limit.

\section{Classical continuum limit of the index density/axial anomaly}

In this section we consider the continuum limit of the index density
$q^U(x)$ (or equivalently, the axial anomaly ${\cal A}(x)=2iq^U(x)$)
for the overlap Dirac operator $D^U$ with $U$ being the lattice transcript 
of a continuum $SU(n)$ gauge field $A$ on ${\bf R}^4$.

Put a hypercubic lattice on ${\bf R}^4$ with lattice spacing $a$.
The space of lattice spinor fields (functions on the lattice sites with
values in ${\bf C}^4\otimes{\bf C}^n$) has the inner product
\be
\la\psi_1,\psi_2\ra=a^4\sum_x\psi_1(x)^*\psi_2(x)
\label{3.1}
\ee
where a contraction of spinor and colour indices is implied. 
For a given lattice
gauge field $U_{\mu}(x)$ the covariant forward (backward) finite difference 
operators $\frac{1}{a}\nabla_{\mu}^+\,$ ($\frac{1}{a}\nabla_{\mu}^-$)
are given by
\be
\nabla_{\mu}^+\psi(x)&=&U_{\mu}(x)\psi(x+ae_{\mu})-\psi(x) \label{3.2} \\
\nabla_{\mu}^-\psi(x)&=&\psi(x)-U_{\mu}(x-ae_{\mu})^{-1}\psi(x-ae_{\mu}) 
\label{3.3}
\ee
where $e_{\mu}$ denotes the unit vector in the positive $\mu$-direction.
Set $\nabla_{\mu}=\frac{1}{2}(\nabla_{\mu}^++\nabla_{\mu}^-)\,$; this 
operator is anti-hermitian with respect to the inner product (\ref{3.1})
since $(\nabla_{\mu}^{\pm})^*=\nabla_{\mu}^{\mp}$. The Wilson--Dirac 
operator is now given by
\be
D_w=\frac{1}{a}\Sn+\frac{r}{2}a(\frac{1}{a^2}\Delta)
\label{3.4}
\ee
where $\Sn=\sum_{\mu}\gamma^{\mu}\nabla_{\mu}$ (the $\gamma^{\mu}$'s are 
taken to be hermitian so $\Sn$ is anti-hermitian), $\Delta=\sum_{\mu}
\nabla_{\mu}^-+\nabla_{\mu}^+=\sum_{\mu}(\nabla_{\mu}^+)^*\nabla_{\mu}^+=
\sum_{\mu}(\nabla_{\mu}^-)^*\nabla_{\mu}^-$ (hermitian, positive) and
$r>0$ is the Wilson parameter. We then have the hermitian operator
\be
H_m=\gamma_5(aD_w-rm)=\gamma_5(\Sn+r({\textstyle \frac{1}{2}}
\Delta-m))
\label{3.5}
\ee
cf. (\ref{2.10}). To define the overlap Dirac operator
\be
D=\frac{1}{a}\Bigl(1+\g5\frac{H}{\sqrt{H^2}}\Bigr)\qquad\qquad\ 
H=H_{m_0}
\label{3.6}
\ee
some restrictions must be made on the lattice fields
(besides excluding the lattice gauge fields for which $H$ has zero-modes).
This can be done in one of the following ways:

\noindent (I) Require $||\psi||<\infty$. Then the lattice spinor fields
form a Hilbert space. Since $H$ is bounded (triangle inequalities give 
$||H||\,\le\,8+8r+rm_0$), $D$ can be defined via the spectral theory
for bounded operators on Hilbert space (see, e.g., \cite{Simon}). 
In this setting no restriction on the lattice gauge field $U_{\mu}(x)$
is required (besides the requirement that $H$ has no zero-modes).
However, the nullspace $\ker{}D$
need not be finite-dimensional in general, so $\index\,D$ is not defined
in general: we can have $\index\,D=\infty-\infty$. 

\noindent (II) Require $\psi(x)$ and $U_{\mu}(x)$ to be periodic in each
direction with fixed periodicity length $L$.
This is equivalent to the finite volume 4-torus setting 
with topologically trivial gauge fields. The space of such lattice spinor 
fields is finite-dimensional, and $H$ leaves this space invariant, 
so $D$ can be defined via the usual spectral
theory and has a well-defined finite index.

\noindent (III) More generally, require $\psi(x)$ and $U_{\mu}(x)$ to be
related at opposite boundaries of a fundamental domain by a gauge 
transformation. This is equivalent to the general finite volume 4-torus 
setting where the gauge fields may be topologically non-trivial. (This was
the setting considered in \cite{Taiwan}.) The space of such lattice
spinor fields is finite-dimensional and $D$ is again well-defined with 
finite index as in (II). 

\noindent (IV) Restrict the lattice to a finite hyper-cubic region
and require that $\psi(x)$ is periodic (or anti-periodic) at
the boundary and $U_{\mu}(x)$ is trivial at the boundary. For such $U$ 
the operator $H$ leaves the (finite-dimensional) space of these spinor fields 
invariant, so $D$ is defined in the usual way and
has well-defined finite index. This setting is a variant
of (II); the difference is that we do not hold the edge length fixed when
taking the (classical) continuum limit. Rather, the limit is taken by
first taking the volume to infinity with fixed spacing $a$, and then
taking $a\to0$ (cf. \S5).

In this paper we focus on the settings (I) and (IV). 
The finite volume torus setting (II)--(III) is treated
in a separate paper \cite{DA(torus)}.

In the following
the density $\O(x,y)$ of an operator $\O$ on the space of lattice spinor
fields is defined through $\O\psi(x)=a^4\sum_y\O(x,y)\psi(y)$. Equivalently,
$\O(x,y)=\frac{1}{a^4}\la\frac{\delta_x}{a^2},\O\frac{\delta_y}{a^2}\ra$
where $\delta_x(y)=\delta_{xy}$.
Note that $\{\frac{\delta_x}{a^2}\}$ is an orthonormal basis for the Hilbert
space of lattice scalar fields in the infinite volume setting (I).
$\O(x,y)$ is a linear operator on 
${\bf C}^4\otimes{\bf C}^n$, and is well-defined when $\O$ is a bounded
operator on the Hilbert space of lattice spinor fields with $||\psi||<\infty$.
The Cauchy--Schwarz inequality gives 
\be
||\O(x,y)||\le\frac{1}{a^4}||\O||\,.
\label{3.6.5}
\ee
In particular, from (\ref{2.7}) and (\ref{2.9}) we see that the 
index density is well-defined:
\be
q^U(x)=-\frac{a}{2}\tr(\g5{}D^U(x,x))=
-\frac{1}{2}\tr\Bigl(\frac{H}{\sqrt{H^2}}(x,x)\Bigr)  
\label{3.7}
\ee
Our interest is in the classical continuum limit of this quantity, i.e
the $a\to0$ limit when $U_{\mu}(x)$ 
is the lattice transcript of a continuum gauge field $A_{\mu}(x)\,$:
\be
U_{\mu}(x)&=&T\,\exp\Bigl(\,\int_0^1aA_{\mu}(x+(1-\tau)ae_{\mu})\,d\tau\Bigr)
\nonumber \\
&=&\sum_{n=0}^{\infty}a^n\int_{0\le\tau_1\le\cdots\le\tau_n\le1}
d\tau_n\cdots{}d\tau_1\,A_{\mu}(x,\tau_n)\cdots{}A_{\mu}(x,\tau_1)
\label{3.8}
\ee
where $A_{\mu}(x,\tau)=A_{\mu}(x+(1-\tau)ae_{\mu})$ and for simplicity
the coupling constant has been set to unity.
The technical setup within which the $a\to0$ limits are taken is as follows.
We assume an infinite collection of hyper-cubic lattices on ${\bf R}^4$
has been specified, parameterised by the lattice spacing $a$, 
with the following properties: (i) For each $\e>0$ there are only finitely 
many lattices with $a\ge\e$. (ii) If $a_1$ and $a_2$ are admissible
lattice spacings and $a_2<a_1$ then the lattice parameterised by
$a_2$ is a subdivision of the one parameterised by $a_1$, i.e.
the sites of the latter are contained in the set of sites of the former.
(E.g. for each $a>0$ we have the lattice with sites  $a{\bf Z}^4$ and 
can use the collection of these lattices parameterised by
$a=a_0,a_0/2,\dots,a_0/2^p,\dots$ for some $a_0>0$.)
The property (ii) implies that if $x\in{\bf R}^4$ is a lattice site
in some lattice with spacing $a$ then it is also a lattice site
in all the other lattices with spacing $a'<a$. 
In the following, in statements concerning $a\to0$ limits 
the variable $x$ always denotes such a point in ${\bf R}^4$; it is fixed 
in ${\bf R}^4$ and does not change as we go from one lattice to another.

To consider the $a\to0$ limit we need the overlap Dirac operator 
$D^U=\frac{1}{a}(1+\g5\frac{H}{\sqrt{H^2}})$ to be 
well-defined for small $a$, i.e. $H$ should not have zero-modes when $a$
is sufficiently small. For technical reasons we will furthermore require
that $H^2$ has a strictly positive $a$-independent lower bound: $H^2>b>0$ 
for sufficiently small $a$. 
The existence of such a bound follows from the results of \cite{local}
(see also \cite{Neu(bound)}). It was shown there that when 
$||1-U(p)||<\e$ for all lattice plaquettes $p$, where $U(p)$ is the product
of the link variables $U_{\mu}(x)$ around $p$, then there is a lower bound
$H^2>b$, depending only on $\e$ and $m_0$, such that for fixed 
$m_0\in(0,2)\,$ $b>0$ when $\e$ is sufficiently small. This result 
generalises to the case of arbitrary $m_0\not\in\{0,2,4,6,8\}$
\cite{DA(bound)}. In the present case, where the link variable is given by 
(\ref{3.8}), the plaquette variable has the standard expansion
\be
1-U(p_{x,\mu,\nu})=a^2F_{\mu\nu}(x)+O(a^3)(x)
\label{3.9}
\ee
and hence
\be
||1-U(p)||\;\sim\;O(a^2)
\label{3.10}
\ee
Strictly speaking this bound requires certain conditions on $A_{\mu}(x)$.
We will discuss these below, but for the moment we proceed under the 
assumption that (\ref{3.10}) is valid, 
i.e. that there exists a finite $K$ independent
of $p$ and small $a$ such that $||1-U(p)||<a^2K$. Then, for arbitrary 
$\e>0$ we have $||1-U(p)||<\e$ for all sufficiently small $a$, and hence
an $a$-independent lower bound $H^2>b>0$ is guaranteed.

To evaluate the continuum limit 
we use an integral representation for $1/\sqrt{H^2}$ to expand it in a 
power series as follows.\footnote{This expansion was used in Kerler's paper  
\cite{K} where it was presented as new.
In fact it had already been given by the present author in \cite{Taiwan}.} 
We first decompose
\be
H^2=L-V
\label{3.11}
\ee
where
\be
L&=&-\nabla^2+r^2({\textstyle \frac{1}{2}}\Delta-m)^2 \label{3.12} \\
V&=&=\;r\frac{1}{2}\gamma_{\mu}V_{\mu}+\frac{1}{4}
[\gamma_{\mu},\gamma_{\nu}]V_{\mu\nu} \label{3.13}
\ee
with
\be
V_{\mu}&=&\frac{1}{2}[(\nabla_{\mu}^++\nabla_{\mu}^-)\,,
\sum_{\nu}(\nabla_{\nu}^--\nabla_{\nu}^+)]
\label{3.14} \\
V_{\mu\nu}&=&\frac{1}{4}\lb(\nabla_{\mu}^++\nabla_{\mu}^-)\,,
(\nabla_{\nu}^++\nabla_{\nu}^-)\rb
\label{3.15} 
\ee
As pointed out in \cite{local}, the norms of the commutators of the
$\nabla_{\mu}^{\pm}$'s are bounded by $max_p||1-U(p)||$. This has the 
following consequences: First, by (\ref{3.10})
\be
||V||\;\sim\;O(a^2)\,.
\label{3.16}
\ee
Furthermore,
\be
V(x,y)=0\quad\mbox{in regions where the gauge field is pure gauge.}
\label{3.17}
\ee
The former implies that
for small $a$ we have $||V||<b/2$ where $b>0$ is the above-mentioned lower
bound on $H^2$. This in turn implies
the lower bound $L>b/2>0$ for the positive operator $L$ in (\ref{3.12}).
Thus for sufficiently small $a$ the operator
$L$ is invertible, $||L^{-1}||\,||V||<1$,
and we can make the expansion
\be
\frac{H}{\sqrt{H^2}}
&=&H\int_{-\infty}^{\infty}\frac{d\sigma}{\pi}\,
\frac{1}{H^2+\sigma^2} \nonumber \\
&=&H\int_{-\infty}^{\infty}\frac{d\sigma}{\pi}\,
\Bigl(\,\frac{1}{1-(L+\sigma^2)^{-1}V}\Bigr) 
\Bigl(\,\frac{1}{L+\sigma^2}\Bigr)
\nonumber \\
&=&H\int_{-\infty}^{\infty}\frac{d\sigma}{\pi}\,
\sum_{k=0}^{\infty}(G_{\sigma}V)^kG_{\sigma}\,.
\label{3.18}
\ee
where $G_{\sigma}:=(L+\sigma^2)^{-1}$. Note that the $\gamma$-matrices
in (\ref{3.11}) are all contained in $V$. Since the trace of $\g5$
times a product of less than 4 $\gamma$-matrices vanishes, the 
$k=0$ and $k=1$ terms in (\ref{3.18}) give vanishing contribution to
$q^U(x)$. On the other hand, the terms with $k\ge3$ satisfy the following 
bound:
\be
&&\Big|\Big|\int_{-\infty}^{\infty}\frac{d\sigma}{\pi}\,
\sum_{k=3}^{\infty}\lb{}H(G_{\sigma}V)^kG_{\sigma}\rb(x,x)\,\Big|\Big|
\nonumber \\
&&\qquad\le\;\frac{1}{a^4}||H||
\int_{-\infty}^{\infty}\frac{d\sigma}{\pi}\,
\sum_{k=3}^{\infty}||G_{\sigma}||^{k+1}||V||^k \nonumber \\
&&\qquad\le\;a^2K^3||H||\left\lb
\int_{-\infty}^{\infty}\frac{d\sigma}{\pi}\,\frac{1}{(b/2+\sigma^2)^4}
\right\rb\,\sum_{k=0}^{\infty}\Bigl(\frac{2}{b}a^2K\Bigr)^k
\label{3.19}
\ee
where we have used (\ref{3.6.5}), (\ref{3.10}) and the bounds 
$G_{\sigma}<(b/2+\sigma^2)^{-1}\le2/b$.
This is $O(a^2)$ since the integral
and sum are finite and remain so in the $a\to0$ limit. Hence only the $k=2$
term in (\ref{3.18}) contributes in the $a\to0$ limit of the index density
(\ref{3.7}), i.e.
\be
q^U(x)=q_2^U(x)+O(a^2)(x)
\label{3.20}
\ee
where
\be
q_2^U(x)=-\frac{1}{2}\int_{-\infty}^{\infty}\frac{d\sigma}{\pi}\,
\tr\lb{}HG_{\sigma}VG_{\sigma}VG_{\sigma}\rb(x,x)\,.
\label{3.21}
\ee
For lattice operators $\O$ which are polynomials in $\nabla_{\mu}^{\pm}$
we denote by $\O^{(0)}$ the operator obtained by setting $U=1$ in
(\ref{3.2})--(\ref{3.3}). Expanding $U_{\mu}(x)$ in powers of $a$
via (\ref{3.8}) gives 
$||H-H^{(0)}||\sim{}O(a)$ and $||L-L^{(0)}||\sim{}O(a)$. (The rigorous
justification of this again requires certain conditions on $A_{\mu}(x)\,$,
to be discussed below.) The latter implies
$||G_{\sigma}-G_{\sigma}^{(0)}||\sim{}O(a)$. This can be seen in various
ways, e.g. as in \cite[(v4)]{DA}, or more simply by noting that
$G_{\sigma}-G_{\sigma}^{(0)}=G_{\sigma}^{(0)}(L^{(0)}-L)G_{\sigma}\,$;
this gives $||G_{\sigma}-G_{\sigma}^{(0)}||\sim{}O(a)$
due to the above-noted upper bound $G_{\sigma},G_{\sigma}^{(0)}<2/b$
which holds for sufficiently small $a$.
This allows us to replace $H$ and $G_{\sigma}$ by $H^{(0)}$ and 
$G_{\sigma}^{(0)}$ in (\ref{3.21}) at the expense of an $O(a)(x)$ term.
Furthermore, since $||V||\sim{}O(a^2)$ we have
$||[L^{(0)},V]||\sim{}O(a^3)$,
which leads to $||[G_{\sigma}^{(0)},V]||\sim{}O(a^3)$ as follows:
The bound $||\nabla_{\mu}^{\pm}||\le2$ and triangle inequalities
lead to an $a$-independent upper bound $L<c$ which allows to expand
\be
G_{\sigma}=\Bigl(\frac{1}{c+\sigma^2}\Bigr)
\Bigl(\frac{1}{1-\frac{c-L}{c+\sigma}}\Bigr)
=\frac{1}{c+\sigma^2}\sum_{m=0}^{\infty}\Bigl(\frac{c-L}{c+\sigma^2}\Bigr)^m
\label{3.21a}
\ee
Now, since 
\be
||[(c-L^{(0)})^m,V]||\;\le\;m||[L^{(0)},V]||\cdot||c-L||^{m-1}
\;\le\;m(a^3K)(c-b/2)^{m-1}
\label{3.21b}
\ee
we get
\be
||[G_{\sigma}^{(0)},V]||\;\le\;\frac{a^3K}{c^2}\sum_{m=0}^{\infty}(m+1)
\Bigl(\frac{c-b/2}{c}\Bigr)^m
\label{3.21c}
\ee
and this is $\sim{}O(a^3)$ since the sum converges (since $0<b/2<c$).
Taking this into account in (\ref{3.21}), it follows from (\ref{3.20}) that
\be
q(x)&=&-\frac{1}{2}\int_{-\infty}^{\infty}\frac{d\sigma}{\pi}\,
\tr\lb{}H^{(0)}V^2(G_{\sigma}^{(0)})^3\rb(x,x)+O(a)(x) \nonumber \\
&=&-\frac{1}{2}\tr\Big\lb{}H^{(0)}V^2
\int_{-\infty}^{\infty}\frac{d\sigma}{\pi}\,\frac{1}{(L^{(0)}+\sigma^2)^3}
\Big\rb(x,x)+O(a)(x) \nonumber \\
&=&\frac{-3}{16}\tr\lb{}H^{(0)}V^2(L^{(0)})^{-5/2}\rb(x,x)+O(a)(x)
\label{3.22}
\ee
Evaluating the trace over spinor indices we find 
\be
q^U(x)&=&\frac{-3r}{16}\e_{\mu\nu\rho\sigma}\tr\Big\lb
(-\nabla_{\mu}^{(0)}(V_{\nu}V_{\rho\sigma}+V_{\nu\rho}V_{\sigma})
+({\textstyle \frac{1}{2}}\Delta^{(0)}-m_0)V_{\mu\nu}V_{\rho\sigma})
(L^{(0)})^{-5/2}\Big\rb(x,x) \nonumber \\
&&\qquad+O(a)(x)
\label{3.23}
\ee
where $V_{\mu}$ and $V_{\mu\nu}$ are given by (\ref{3.14})--(\ref{3.15}).
Noting that \cite{local}
\be
\lb\nabla_{\mu}^+,\nabla_{\nu}^+\rb\psi(x)
=(1-U(p_{x,\mu\nu}))U_{\mu}(x)U_{\nu}(x+ae_{\mu})\psi(x+ae_{\mu}+ae_{\nu})
\label{3.24}
\ee
and similar formulae for the other commutators,
calculations with (\ref{3.8}) give
\be
\lb\nabla_{\mu}^{\pm},\nabla_{\nu}^{\pm}\rb\psi(x)
&=&(a^2F_{\mu\nu}(x)+O(a^3)(x))\psi(x\pm{}ae_{\mu}\pm{}ae_{\nu}) 
\label{3.25} \\
\lb\nabla_{\mu}^{\pm},\nabla_{\nu}^{\mp}\rb\psi(x)
&=&(a^2F_{\mu\nu}(x)+O(a^3)(x))\psi(x\pm{}ae_{\mu}\mp{}ae_{\nu}) \label{3.26}
\ee
These determine the relevant contributions of $V_{\mu}$ and $V_{\mu\nu}$
in (\ref{3.23}).

We now exploit the fact that the delta-function $\delta_x$ on the lattice
sites has the Fourier expansion in plane wave fields:
\be
\delta_x=\int_{-\pi}^{\pi}\frac{d^4k}{(2\pi)^4}\,e^{-ikx/a}\phi_k
\label{3.27}
\ee
where $\phi_k(y):=e^{iky/a}$.
For a general operator $\O$ this leads to 
\be
\O(x,x)&=&\frac{1}{a^4}\la\frac{\delta_x}{a^2},\O\frac{\delta_x}{a^2}\ra
=\frac{1}{a^4}\int_{-\pi}^{\pi}\frac{d^4k}{(2\pi)^4}\,e^{-ikx/a}
\frac{1}{a^4}\la\delta_x,\O\phi_k\ra \nonumber \\
&=&\frac{1}{a^4}\int_{-\pi}^{\pi}\frac{d^4k}{(2\pi)^4}\,e^{-ikx/a}
(\O\phi_k)(x)
\label{3.28}
\ee
In the case where
\be
\O=\e_{\mu\nu\rho\sigma}
(-\nabla_{\mu}^{(0)}(V_{\nu}V_{\rho\sigma}+V_{\nu\rho}V_{\sigma})
+({\textstyle \frac{1}{2}}\Delta^{(0)}-m_0)V_{\mu\nu}V_{\rho\sigma})
(L^{(0)})^{-5/2}
\label{3.29}
\ee
a calculation using (\ref{3.14})--(\ref{3.15}) with 
(\ref{3.25})--(\ref{3.26}) gives\footnote{Essentially the same calculations
were presented as new in \cite{K}. In fact they had already been done
in \cite[(v4)]{DA} and \cite{Taiwan}.}
\be
(\O\phi_k)(x)=32\pi^2a^4\lambda(k;r,m_0)(q^A(x)+O(a)(x))\phi_k(x)
\label{3.30}
\ee
where $q^A(x)$ is the continuum index density (\ref{1}), and
\be
\lambda(k;r,m)=
\frac{\prod_{\nu}\cos{}k_{\nu}\Bigl(-m+
\sum_{\mu}(1-\cos{}k_{\mu})-\sum_{\mu}\frac{\sin^2k_{\mu}}{\cos{}k_{\mu}}
\Bigr)}{\Big\lb\,\sum_{\mu}\sin^2k_{\mu}+
r^2(-m+\sum_{\mu}(1-\cos{}k_{\mu}))^2\Big\rb^{5/2}}
\label{3.31}
\ee
It follows from (\ref{3.23}) and (\ref{3.28}) that
\be
q(x)=I(r,m_0)q^A(x)+O(a)(x)
\label{3.32}
\ee
where
\be
I(r,m)=\frac{-3r}{8\pi^2}\int_{-\pi}^{\pi}d^4k\,\lambda(k;r,m)\,.
\label{3.34}
\ee
The integral $I(r,m)$
is similar to the integral (A.17) of \cite{Seiler}, although the 
exponents and numerical factor are different 
and the parameter $m$ did not appear there.
To evaluate it we exploit the symmetries of the integrand (as in 
\cite{Seiler}) and change variables to $s_{\nu}\equiv\sin{}k_{\nu}$
to write
\begin{eqnarray}
I(r,m)=\sum_{\epsilon_{\mu}=\pm1}\Bigl(\,\prod_{\mu=1}^4
\mbox{sign}(\epsilon_{\mu})\Bigr)I(r,m,\epsilon)
\label{t19.5}
\end{eqnarray}
where
\begin{eqnarray}
I(r,m,\epsilon)
=\frac{-3r}{8\pi^2}\int_{-1}^1d^4s\frac{-m+\seu-\sum_{\mu}\frac{s_{\mu}^2}
{\epsilon_{\mu}\sqrt{1-s_{\mu}^2}}}
{\Big\lb\,s^2+r^2\Bigl(-m+\seu\Bigr)^2\,\Big\rb^{5/2}}
\label{t20}
\end{eqnarray}
This diverges for $m=m_{\epsilon}\equiv\sum_{\mu}(1-\epsilon_{\mu})
\in\{0,2,4,6,8\}$
but is finite for all other values of $m$. It is a constant function of
$m$ with a jump at $m_{\epsilon}\,$; to see this set
$\wt{\Delta}(s,m)=-m+\seu\,$, then
\begin{eqnarray}
\frac{-8\pi^2}{3}\frac{d}{dm}I(r,m,\epsilon)
=-r\int_{-1}^1\frac{d^4s}{(s^2+r^2\wt{\Delta}^2)^{5/2}}
+5r^3\int_{-1}^1d^4s\frac{\wt{\Delta}(1-\sum_{\nu}s_{\nu}
\frac{\partial}{\partial{}s_{\nu}})\wt{\Delta}}
{(s^2+r^2\wt{\Delta}^2)^{7/2}}
\label{t21}
\end{eqnarray}
Inspired by the identity eq. (A.19) of \cite{Seiler} (which was originally
due to Karsten and Smit) we rewrite the second integral as
\begin{eqnarray}
5r\int_{-1}^1d^4s\frac{(1-\frac{1}{2}\sum_{\nu}s_{\nu}
\frac{\partial}{\partial{}s_{\nu}})(s^2+r^2\wt{\Delta}^2)}
{(s^2+r^2\wt{\Delta}^2)^{7/2}} \qquad\qquad & & \nonumber \\
=5r\int_{-1}^1d^4s\Bigl(\,1+{\textstyle \frac{1}{5}}\sum_{\nu}s_{\nu}
{\textstyle \frac{\partial}{\partial{}s_{\nu}}}\,\Bigr)
(s^2+r^2\wt{\Delta}^2)^{-5/2} & &
\label{t21.5}
\end{eqnarray}
Integration by parts now gives the first integral in (\ref{t21}) 
except with the 
opposite sign, so (\ref{t21}) vanishes. $I(r,m)$ can now be 
determined by evaluating $I(r,m,\epsilon)$ 
in the limits $m\to{}m_{\epsilon}$ from above and below.
For $m>m_{\epsilon}\,$, after setting $m=m_{\epsilon}+\td{m}$
and changing variables to $\td{s}_{\nu}=s_{\nu}/\td{m}$ in (\ref{t20}),
we get
\begin{eqnarray}
I(r,m,\epsilon)
&=&\frac{-3r}{8\pi^2}\int_{-1/\td{m}}^{1/\td{m}}d^4\td{s}\,\td{m}^4
\frac{-\td{m}+O(\td{m}^2)}
{\td{m}^5\lb{}\td{s}^2+r^2(-1+O(\td{m}))^2\,\rb^{5/2}} \nonumber \\
&\stackrel{\td{m}\to0_+}{=}&\frac{-3}{8\pi^2}\int_{-\infty}^{\infty}
d^4\td{s}\,\frac{-r}{(\td{s}^2+r^2)^{5/2}}\;=\;1/2 
\label{t22}
\end{eqnarray}
For $m<m_{\epsilon}$ a similar calculation gives $I(r,m,\epsilon)\!=\!-1/2$.
Thus $I(r,m)$ is independent of $r>0$ and
can now be calculated from (\ref{t19.5}), leading to the value
$I(m)$ in table below:\footnote{The evaluation of this integral was given in
the first version \cite[(v1)]{DA} of this paper, and also (in more detail)
in \cite{Suz}.}
\hfill\break
\begin{tabular}{l|c|c|c|c|c|}
          & $0<m<2$ & $2<m<4$ & $4<m<6$ & $6<m<8$ & 
$m\not\in[0,8]$ \\ \cline{2-6}
$I(m)\;$= & $1$     &  $-3$   & $ 3$    &  $-1$   &  $ 0$ \\
\end{tabular}
\be
\label{3.35}
\ee
By (\ref{3.32}) we now have
\be
q^U(x)=I(m_0)q^A(x)+O(a)(x)
\label{3.36}
\ee
Thus for $0<m_0<2$ the index density $q^U(x)$ does indeed have the 
correct classical continuum limit in the infinite volume setting.

The $O(a^p)$
bounds used in the calculations above can be established
by standard calculations when the continuum field $A_{\mu}(x)$ is smooth
and $||A_{\mu}(x)||\,$, $||\partial_{\mu}A_{\nu}(x)||$ and 
$||\partial_{\mu}\partial_{\nu}A_{\rho}(x)||$ are all bounded on 
${\bf R}^4$ (the details can be found in the appendix of \cite{DA(torus)}).
However, the result $q(x)=q^A(x)+O(a)(x)$ (and its generalisation 
(\ref{3.36}) when $m_0\not\in[0,2]$) holds in more general cases as well:
It suffices that 
the plaquette variable of the lattice transcript of $A$ satisfies
$||1-U(p)||<\e$ when the lattice spacing $a$ is sufficiently small,
with $\e$ small enough that a strictly positive lower bound $H^2>b>0$ 
is guaranteed (as discussed earlier). 
Then $D^U$ is local in the gauge field \cite{local}, leading to
\be
|q^U(x)-q^{\td{U}}(x)|\;\sim\;
O({\textstyle \frac{1}{a^4}}\,e^{-\rho/a})
\label{3.37}
\ee 
where $\td{U}$ is the lattice transcript
of a smooth continuum field $\td{A}$ which coincides
with $A$ in a neighbourhood of $x$ and vanishes outside a bounded region
of ${\bf R}^4$. Since such $\td{A}$ and its partial derivatives are
automatically bounded on ${\bf R}^4$, we have 
$q^{\td{U}}(x)=q^A(x)+O(a)(x)$, and this together with (\ref{3.37}) 
gives $q(x)=q^A(x)+O(a)(x)$.

A more detailed justification of the preceding is as follows.
Let $A$ be an arbitrary smooth continuum gauge field on ${\bf R}^4$.
Pick a smooth function $\lambda(y)$ on ${\bf R}^4$ which is equal to 
1 in the box
$\prod_{\mu}[x_{\mu}-de_{\mu},x_{\mu}+de_{\mu}]\subset{\bf R}^4$
($d>0$) and which vanishes outside a bounded region of ${\bf R}^4$.
To prove the claims above, it suffices to establish (\ref{3.37}) for the 
case where $\td{A}_{\mu}(y)=\lambda(y)A_{\mu}(y)$. For this
we exploit the fact \cite{local} that there is a power series expansion
$1/\sqrt{H^2}=\kappa\sum_{k=0}^{\infty}t^kP_k(H^2)$
where $P_k(\cdot)$ is a Legendre polynomial of order $k\,$; 
$||P_k(H^2)||\le1\,$; $t=e^{-\theta}\,$; the constants $\kappa,\theta>0$
depend only on the (strictly positive) lower and upper bounds on $H^2$
\cite{local}. Similarly 
$1/\sqrt{\ttd{H}^2}=\kappa\sum_{k=0}^{\infty}t^kP_k(\ttd{H}^2)$
where $\ttd{H}$ is obtained from $H$ by replacing the lattice gauge field
$U$ (the lattice transcript of $A$) by the lattice transcript $\td{U}$
of $\td{A}$.
(This requires $H^2$ and $\ttd{H}^2$ to have a lower bound $b>0$.
This is guaranteed in both cases when $a$ is sufficiently small:
In the first case we have $||1-U(p)||<\e$ by assumption, while in the second
case $||1-\td{U}(p)||\sim{}O(a^2)$ since $\td{A}_{\mu}(x)$ and its
partial derivatives are automatically bounded on ${\bf R}^4$.)
Since $H$ only couples nearest neighbour sites, $P_k(H^2)$ can only couple
the site $x$ to itself via a site outside of 
$\prod_{\mu}[x_{\mu}-de_{\mu},x_{\mu}+de_{\mu}]\subset{\bf R}^4$
if $k\ge2(d/2a)$. Therefore $[P_k(H^2)](x,x)=[P_k(\ttd{H}^2)](x,x)$
when $k<d/a$, and it follows that
\be
\Big|\Big|\,\frac{1}{\sqrt{H^2}}(x,x)-\frac{1}{\sqrt{\ttd{H}^2}}(x,x)
\,\Big|\Big|
&\le&\kappa\sum_{k\ge{}d/a}^{\infty}t^k\,||\,\lb{}P_k(H^2)\rb(x,x)
-\lb{}P_k(\ttd{H}^2)\rb(x,x)\,|| \nonumber \\
&\le&\kappa\,t^{d/a}\sum_{k=0}^{\infty}t^k\frac{1}{a^4} \nonumber \\
&=&\Bigl(\frac{\kappa}{1-e^{-\theta}}\Bigr)\,\frac{1}{a^4}\,e^{-\theta{}d/a}
\label{3.38}
\ee
This together with the fact that $H$ is ultra-local gives (\ref{3.37}).

The preceding observations allow us to conclude
that $q^U(x)=q^A(x)+O(a)(x)$ in some cases where $A_{\mu}(x)$ diverges
at infinity, or has singularities. An example of the former is a
topologically non-trivial gauge field on the 4-torus: these can be viewed
as gauge fields on ${\bf R}^4$ satisfying a periodicity condition
\be
A_{\mu}(x+Le_{\nu})=\Omega(x,\nu)A_{\mu}(x)\Omega(x,\nu)^{-1}
+\Omega(x,\nu)\partial_{\mu}\Omega(x,\nu)^{-1}
\label{3.39}
\ee
where $\Omega(x,\nu)\,$, $\nu=1,2,3,4\,$, are the $SU(n)$-valued monodromy
fields which specify the principal $SU(n)$ bundle over $T^4$. 
Fields $A_{\mu}(x)$ satisfying (\ref{3.39}) diverge at infinity in the
topologically non-trivial case. However, the requirement 
$||1-U(p)||<\epsilon$ for small $a$ still holds in this case
since (i) the lattice transcript of $A_{\mu}(x)$ satisfies
\be
U_{\mu}(x+Le_{\nu})=\Omega(x,\nu)U_{\mu}(x)\Omega(x+ae_{\nu},\nu)^{-1}
\label{3.40}
\ee
and (ii) $||U_{\mu}(y)||=1$ for all $y,\mu$ since $U_{\mu}(y)$ is unitary.
These imply that if $x_{\rho}'=x_{\rho}+Ln_{\rho}e_{\rho}$ then
$||1-U(p_{x,\mu\nu})||=||1-U(p_{x',\mu\nu})||$ for
for all $n\in{\bf Z}^4$.
It follows that $||1-U(p)||\sim{}O(a^2)$ since this is true for plaquettes 
in the fundamental domain $[0,L]^4\subset{\bf R}^4$ because
$A_{\mu}(x)$ and its partial derivatives are automatically bounded in 
$[0,L]^4$.

An example where the requirement $||1-U(p)||<\e$ for small $a$ 
is satisfied when
$A_{\mu}(x)$ has a singularity is the following: Consider a gauge field
in a singular gauge, and such that $A_{\mu}(x)$ is pure gauge in a 
neighbourhood of the singularity and vanishes outside a bounded region of 
${\bf R}^4$. (Examples of topologically non-trivial gauge fields of
this type are readily constructed, cf. \S5.) We choose the lattices on
${\bf R}^4$ such that the singularity of $A_{\mu}(x)$ never lies on
a link of any lattice (cf. \S5). Then the lattice transcript $U_{\mu}(x)$
is well-defined for all lattices. For lattice plaquettes contained
in the neighbourhood of the singularity we have $U(p)=1$ since
$A_{\mu}(x)$ is pure gauge. On the other hand, $A_{\mu}(x)$ and its 
partial derivatives are bounded outside the neighbourhood of the singularity
(since the field vanishes outside a bounded region). 
Hence $||1-U(p)||\sim{}O(a^2)$.

\section{Aspects of $\index\,D^U$ and its continuum limit in the infinite
volume setting}

In contrast to $q^U(x)$, 
the index of $D^U$ is a problematic quantity in the infinite volume setting.
It is not well-defined a priori by
\be
\index\,D^U=\Tr\Bigl(\g5\Big|_{\ker{}D^U}\Bigr)
\label{4.1}
\ee
since the null-space $\ker{}D^U$ may be infinite-dimensional.
The same is true for the index of the 
continuum Dirac operator $\sd^A$ for general gauge field $A_{\mu}(x)$
on ${\bf R}^4$. In the latter case
the index exists and is equal (by the index theorem) to 
$Q(A)=\int_{{\bf R}^4}d^4x\,q^A(x)$ provided $A_{\mu}(x)$ 
converges sufficiently
quickly to a pure gauge configuration at infinity. A natural conjecture
in the lattice setting is therefore that $\index\,D^U\,$, given by 
(\ref{4.1}), exists and is equal to $a^4\sum_xq^U(x)$ when the lattice
gauge field $U_{\mu}(x)$ converges sufficiently quickly to a pure
gauge configuration at infinity. To prove this is a challenging problem 
though, and in this section we will only make some first steps towards it.
We investigate what happens if $q^U(x)$ is replaced by
$a^4\sum_xq^U(x)$ in the continuum limit calculation of the preceding 
section; this will indicate how a situation where $U_{\mu}(x)$
is pure gauge at infinity can result in the infinite sum $\sum_xq^U(x)$
being convergent and $a^4\sum_xq^U(x)\to{}Q(A)$ in the classical
continuum limit. However we do not show that $\index\,D^U$ exists
and is equal to $a^4\sum_xq^U(x)$ in this case; this remains as a problem
for future work.

Using the expansion (\ref{3.18}) we get
\be
a^4\sum_xq^U(x)=-\frac{1}{2}a^4\sum_x
\int_{-\infty}^{\infty}\frac{d\sigma}{\pi}\,\sum_{k=0}^{\infty}
\tr\lb{}H(G_{\sigma}V)^kG_{\sigma}\rb(x,x)
\label{4.2}
\ee
As before, the $k=0$ and $k=1$ terms give vanishing contribution.
Obviously, to get $a^4\sum_xq^U(x)\to\int_{{\bf R}^4}d^4x\,q^A(x)=Q(A)$
we will need the part with $k\ge3$ to vanish in the $a\to0$ limit as in \S3.
This cannot be expected to happen in general since the infinite sum
$\sum_x$ might not even be convergent for $k\ge3$ part of (\ref{4.2}).
However, as we will now show, the sum over lattice sites does converge
when $U_{\mu}(x)$ is pure gauge outside a bounded region ${\cal R}$ of 
${\bf R}^4$. Furthermore, we will see that 
when $U_{\mu}(x)$ is the lattice transcript
of a smooth continuum field $A_{\mu}(x)$ which is pure gauge outside of
${\cal R}$ the $k\ge3$ part in (\ref{4.2}) does indeed vanish in the 
$a\to0$ limit and we get $a^4\sum_xq^U(x)\to{}Q(A)$.
We can assume that $\R$ is a 4-dimensional box $[-L,L]^4$. Let $\R_d$
denote the larger box $[-L-d,L+d]^4$ ($d>0$). The strategy is to split
up the sum over lattice sites in (\ref{4.2}) into a finite sum
$\sum_{x\in\R_d}(\cdots)$ and an infinite sum 
$\sum_{x\in{\bf R}^4-\R_d}(\cdots)$ where the sums are over the lattice sites
in $\R_d$ and ${\bf R}^4-\R_d$ respectively. The first sum is easy to deal 
with after noting that 
$a^4\sum_{x\in\R_d}(\cdots)\to\int_{\R_d}d^4x\,(\cdots)$
for $a\to0$ and using the fact that the volume $V(\R_d)$ is finite.
We will deal with the second sum by exploiting the facts that
$G_{\sigma}(z,x)$ is (exponentially) local and $V(y,z)$ is ultra-local
and vanishes in regions where $U$ is pure gauge.

By the same calculations as in (\ref{3.19}) we see that the $k\ge{}p$ part
of the summand in (\ref{4.2}) has a bound
\be
\Big|\,\int_{-\infty}^{\infty}\frac{d\sigma}{\pi}\,\sum_{k=p}^{\infty}
\tr\lb{}H(G_{\sigma}V)^kG_{\sigma}\rb(x,x)\,\Big|
\;\le\;a^{2(p-2)}K_p
\label{4.3}
\ee
($p\ge2$) for some constant $K_p$ independent of $a$ and $x$.
Choose an $\e>0$, then for sufficiently small $a$ we have 
$a^4\sum_{x\in\R_d}1\le{}V(\R_d)+\e\,\equiv\,V(\R_d)_{\e}$ where 
$V(\R_d)=(2(L+d))^4$ is the volume of $R_d$.
It follows that, for small $a$,
\be
\Big|\,a^4\sum_{x\in\R_d}
\int_{-\infty}^{\infty}\frac{d\sigma}{\pi}\,\sum_{k=p}^{\infty}
\tr\lb{}H(G_{\sigma}V)^kG_{\sigma}\rb(x,x)\,\Big|
\;\le\;a^{2(p-2)}K_pV(\R_d)_{\e}
\label{4.4}
\ee
This vanishes in the $a\to0$ limit for $p=3$, and does not diverge
for $p=2$. Hence when the sum over $x$ in (\ref{4.2}) is restricted to
the lattice sites in $\R_d$ the $k\ge3$ part vanishes in the $a\to0$
limit as required, and the $k=2$ part remains finite.
We now consider the summand in (\ref{4.2}) for
$x$ outside of $\R_d$. We use 
\be
\lb{}H(G_{\sigma}V)^kG_{\sigma}\rb(x,x)
=a^8\sum_{y,z}\lb{}H(G_{\sigma}V)^{k-1}G_{\sigma}\rb(x,y)
V(y,z)G_{\sigma}(z,x)
\label{4.5}
\ee
By (\ref{3.17}) $V(y,z)=0$ for $y,z\not\in\R$ so the sum over $y$ and $z$ 
in (\ref{4.5}) can be restricted to the lattice sites in $\R$.
We now apply a version of the locality argument of \cite{local} to
$G_{\sigma}\,$: For small $a$ we have $a$-independent bounds
$0<b/2<L<c$ and get a power series expansion of $G_{\sigma}$ as in
\ref{3.21a},
\be
G_{\sigma}
=\frac{1}{c+\sigma^2}\sum_{k=0}^{\infty}\Bigl(\frac{c-L}{c+\sigma^2}
\Bigr)^k
\label{4.6}
\ee
which converges since $||\frac{c-L}{c+\sigma}||<\frac{c-b/2}{c+\sigma^2}
<\frac{c-b/2}{c}\equiv{}t<1$. Since $L$ only couples nearest neighbour and 
next to nearest neighbour sites, $L^k(x,y)=0$ when 
$\sum_{\mu}|x_{\mu}-y_{\mu}|/a>2k$. It follows that 
\be
||G_{\sigma}(x,y)||&\le&
\frac{1}{c+\sigma^2}\sum_{k\ge\frac{1}{2}\sum_{\mu}|x_{\mu}-y_{\mu}|/a}
\Big|\Big|\,\Bigl(\frac{c-L}{c+\sigma^2}\Bigr)^k\,\Big|\Big| 
\;\le\;||G_{\sigma}||\,t^{\Bigl(\frac{1}{2}\sum_{\mu}|x_{\mu}-y_{\mu}|/a\Bigr)}
\sum_{k=0}^{\infty}t^k/a^4 \nonumber \\
&\le&||G_{\sigma}||\,\frac{\td{\kappa}}{a^4}\,\exp\Bigl(-\frac{\theta}{2}
\sum_{\mu}|x_{\mu}-y_{\mu}|/a\Bigr)
\label{4.7}
\ee
where $t=e^{-\theta}=\frac{c-b/2}{c}$ and $\td{\kappa}=1/(1-t)$. 
Applying this to $G_{\sigma}(z,x)$ in (\ref{4.5}) gives
\be
||G_{\sigma}(z,x)||\;\le\;||G_{\sigma}||\,\frac{\td{\kappa}}{a^4}
\,\exp\Bigl(-\frac{\theta}{2}\sum_{\mu}(|x_{\mu}|-L)/a\Bigr)
\label{4.8}
\ee
since $|x_{\mu}-z_{\mu}|>|x_{\mu}|-L$ when $z\in\R$ and
$x$ is outside of $\R_d$. 
This leads to a bound on the part of (\ref{4.2}) where the sum
is restricted to the lattice sites in ${\bf R}^4-\R_d\,$:  
\be
&&\Big|\,a^4\sum_{x\in{\bf R}^4-\R_d}
\int_{-\infty}^{\infty}\frac{d\sigma}{\pi}\,\sum_{k=0}^{\infty}
\tr\lb{}H(G_{\sigma}V)^kG_{\sigma}\rb(x,x)\,\Big| \nonumber \\
&&\quad=\Big|\,a^4\sum_{x\in{\bf R}^4-\R_d}a^8\sum_{y,z\in\R}
\int_{-\infty}^{\infty}\frac{d\sigma}{\pi}\,\sum_{k=2}^{\infty}
\tr\lb{}H(G_{\sigma}V)^{k-1}G_{\sigma}\rb(x,y)V(y,z)G_{\sigma}(z,x)\,\Big|
\nonumber \\
&&\quad\le\;
a^4\sum_{x\in{\bf R}^4-\R_d}\sum_{y,z\in\R}
\int_{-\infty}^{\infty}\frac{d\sigma}{\pi}\,\sum_{k=2}^{\infty}
4n||H(G_{\sigma}V)^{k-1}G_{\sigma}||\,||V||\,||G_{\sigma}(z,x)||
\nonumber \\
&&\quad\le\;\left\lb\frac{1}{a^4}
\int_{-\infty}^{\infty}\frac{d\sigma}{\pi}\,\sum_{k=2}^{\infty}
4n||H(G_{\sigma}V)^kG_{\sigma}||\right\rb
\,\left(\,a^8\sum_{y,z\in\R}1\right)\ \times \nonumber \\
&&\quad\quad\times\ a^4\sum_{x\in{\bf R}^4-\R_d}\frac{\td{\kappa}}{a^4}
\,\exp\Bigl(-\frac{\theta}{2}\sum_{\mu}(|x_{\mu}|-L)/a\Bigr)
\label{4.9}
\ee
Since $||V||\sim{}O(a^2)$ the integral is $\sim{}O(a^4)$. For small $a$
we have $a^8\sum_{y,z\in\R}1\le(V(\R)_{\e})^2$. 
Finally, the sum is bounded by
\be
\frac{\td{\kappa}}{a^8}\prod_{\mu}\left\lb\,2\int_{L+d}^{\infty}
\exp\Bigl(-\frac{\theta}{2a}(t_{\mu}-L)\Bigr)\,dt_{\mu}\,\right\rb
=\td{\kappa}\Bigl(\frac{2}{a\theta}\Bigr)^4e^{-2\theta{}d/a}
\label{4.10}
\ee
Thus (\ref{4.9}) vanishes as $O(\frac{1}{a^4}\,e^{-\rho/a})$ for $a\to0$.
This completes
the demonstration that the sum over lattice sites in (\ref{4.2}) is
convergent when $U$ is pure gauge outside a bounded region,
and that the $k\ge3$ part vanishes in the $a\to0$
limit when $U$ is the lattice transcript of a continuum field $A$ which
is pure gauge outside a bounded region. Furthermore, the preceding shows
that in this case the non-vanishing contribution to $a^4\sum_xq^U(x)$
in the $a\to0$ limit comes from $a^4\sum_{x\in\R_d}q_2^U(x)$ where
the subscript ``2'' refers to the $k=2$ part of (\ref{4.2}).
It is a straightforward technical exercise to show that this 
reduces to $\int_{\R_d}d^4x\,q^A(x)$ in the $a\to0$ limit
along the same lines as the argument for $q^U(x)\to{}q^A(x)$
in \S3 (we omit the details). By Stokes theorem this reduces to the integral
of the Chern-Simons term over the boundary $\partial\R_d$, 
which calculates the winding number of the map from 
$\partial\R_d\cong{}S^3$ to $SU(n)$ corresponding to the pure gauge 
configuration, and this number is precisely $Q(A)$. 

Clearly the preceding arguments can be generalised from the case where
$U$ and $A$ are pure gauge outside a bounded region of ${\bf R}^4$
to cases where the fields converge sufficiently quickly to pure gauge
configurations at infinity. To determine a precise criterion
for what is ``sufficiently quickly'' is a non-trivial problem though, and is
beyond the scope of this paper.
In the continuum the criterion is
that $F_{\mu\nu}(x)$ should vanish quickly enough at infinity to be 
square-integrable \cite{Uhlenbeck}. We speculate that a similar condition 
will suffice in the lattice setting.

\section{Classical continuum limit of $\index\,D^U$ starting in a finite 
volume lattice setting and taking an infinite volume limit}

In this section we consider $\index\,D^U$ in the lattice setting (IV) of \S3.
The volume is finite, hence the space of lattice spinor fields is
finite-dimensional and $\index\,D^U$ is well-defined from the beginning.
The boundary condition on the lattice gauge field is 
that it is trivial at the boundary (i.e. the Dirichlet
condition). Then the covariant finite difference operators 
$\nabla_{\mu}^{\pm}$ map both the spaces of periodic and anti-periodic
lattice spinor fields to themselves, and in both cases satisfy
$(\nabla_{\mu}^{\pm})^*=-\nabla_{\mu}^{\mp}\,$, so $H$ is hermitian and
$D^U$ is well-defined. We can therefore take either periodic or anti-periodic
boundary conditions on the lattice spinor fields. 
We consider the case where $U$ is the lattice 
transcript of a continuum gauge field $A_{\mu}(x)$ in a singular 
gauge and vanishing outside a bounded region of ${\bf R}^4$. The singularity
allows $A$ to be topologically non-trivial (i.e. $Q(A)\ne0$) while
the latter requirement ensures that $U_{\mu}(x)$ is trivial outside
of a bounded region of ${\bf R}^4$ and therefore satisfies the Dirichlet
boundary condition when the volume is sufficiently large. 
We will show that $\index\,D^U$ reduces to the continuum topological
charge $Q(A)$ in the classical continuum limit defined by first taking
the infinite volume limit and then taking $a\to0$.
Since $\index\,D^U$ and $Q(A)$ are both integers, it then follows that
$\index\,D^U=Q(A)$ for all sufficiently large lattices with sufficiently 
small lattice spacings.
This demonstrates analytically that $\index\,D^U$ is able to capture
the topological data of the continuum field $A$ (complementing 
previous numerical results which we discuss later).
The present setting is arguably the simplest in which such a result can be
analytically demonstrated. 
The same result holds in the finite volume torus
setting \cite{DA(torus)}; however, the argument is less simple
than what follows.\footnote{In fact the result derived
in this section can also be derived quite straightforwardly as an 
application of the result for the finite volume torus case 
\cite{L(private)}.}

Let $A_{\mu}(x)$ be an $SU(n)$ gauge field on ${\bf R}^4$ which may be
singular at the origin but is smooth everywhere else and vanishes
outside a bounded region. Furthermore we require that $A$ is pure gauge
in a neighbourhood of the origin. Examples of such fields are readily
obtained as follows. 
Take a smooth map $\phi:S^3\to{}SU(n)$
with degree $Q$ and define $\td{\phi}:{\bf R}^4-\{0\}\to{}SU(n)$
by $\td{\phi}(y,t)=\phi(y)\,$, $y\in{}S^3\,$, $t\in{\bf R_+}\,$,
where we are identifying ${\bf R}^4-\{0\}$ with $S^3\times{\bf R_+}$
in the obvious way.
Choose a smooth real function $\lambda(x)$ on ${\bf R}^4$ equal to 1 
in a neighbourhood of the origin and vanishing outside a bounded region.
Then the field
\be
A_{\mu}(x)=\lambda(x)\td{\phi}(x)\partial_{\mu}\td{\phi}(x)^{-1}
\label{5.1}
\ee
has topological charge $Q$, is singular at the origin, pure gauge in a
neighbourhood of the origin and vanishes outside the bounded region.

We take the finite hyper-cubic lattice in ${\bf R}^4$ to have spacing $a$
and sites
\be
\lbrace{}x=a(n_1-1/2\,,n_2-1/2\,,n_3-1/2\,,n_4-1/2)\}\;|\;n\in{\bf Z}^4\; 
-N<n_{\mu}\le{}N\;\rbrace
\label{5.2}
\ee 
Note that the origin $0\in{\bf R}^4$ never lies on a
lattice link, so the lattice transcript $U_{\mu}(x)$ of $A_{\mu}(x)$ 
(given by (\ref{3.8})) is well-defined for all $a$ and $N$.

Our aim in this section is to show that, with either periodic or anti-periodic
boundary conditions on the lattice spinor fields,
\be
\lim_{a\to0}\;\lim_{N\to\infty}\;\index\,D^U=I(m_0)Q(A)
\label{5.3}
\ee
where $I(m_0)$ is given by the table (\ref{3.35}) (in particular $I(m_0)=1$
for the physical values $0<m_0<2$).
The techniques used to derive this have already been developed in the 
earlier parts of this paper, so we will be economical with the details in
the following.

\noindent {\it Proof of (\ref{5.3}).} 
We start from the formula
\be
\index\,D^U=a^4\sum_xq^U(x)\quad,\quad\ 
q^U(x)=-\frac{1}{2}\tr\frac{H}{\sqrt{H^2}}(x,x)
\label{5.4}
\ee
where the sum is over the lattice sites (\ref{5.2}).
Choose $L$ sufficiently large so that $A_{\mu}(x)$ vanishes for all $x$ 
outside of $\R=[-L,L]^4$. Choose $d>0$ and set $\R_d=[-L-d,L+d]^4$ as
before. 
By the argument at the end of \S3 we have $||1-U(p)||\sim{}O(a^2)$.
This leads to a lower bound $H^2>b>0$ for small $a$, with $b$ independent
of both $a$ and $N$. Therefore the previous bounds such as 
$||V||\le{}a^2K$ and $L>b/2$ continue to hold in the present setting
and can be chosen independent of $N$. We henceforth restrict to the 
(sufficiently small) lattice spacings $a$ for which these bounds are
satisfied. Furthermore, for given $a$ we restrict to the (sufficiently large)
$N$'s for which $aN>L+d$ so that the lattice (specified by (\ref{5.2}))
covers $\R_d$. The $N\to\infty$ limit of the part of (\ref{5.4}) where the 
sum is restricted to sites outside of $\R_d$ has the bound
\be
\lim_{N\to\infty}\;\Big|\,a^4\sum_{x\not\in\R_d}q^U(x)\,\Big|
\;\le\;a^4\sum_{x\in{\bf R}^4-\R_d}|q^U(x)|
\label{5.5}
\ee
where the sum on the right-hand side is over all lattice sites outside
of ${\bf R}^4-\R_d$ in the extension of the lattice to the whole of
${\bf R}^4$. This vanishes in the $a\to0$ limit by a similar argument to
the one based on (\ref{4.9})--(\ref{4.10}) in the previous section.
Thus in deriving (\ref{5.3}) it suffices to restrict the sum in (\ref{5.4}) 
to the lattice sites in $\R_d$. 

Now choose $L'>0$ small enough that $[-L',L']^4$ is contained within
the neighbourhood of the origin in ${\bf R}^4$ where $A_{\mu}(x)$ is
pure gauge. Choose $e>0$ so that $e<L'$ (e.g. $e=L'/2$) and
set $\R_e=[-L'+e,L'-e]^4$. By a straightforward adaptation of the argument
based on (\ref{4.9})--(\ref{4.10}) we find that the part of (\ref{5.4})
where the sum is restricted to the lattice sites in $\R_e$ vanishes
in the $a\to0$ limit after taking $N\to\infty$.  
Thus in deriving (\ref{5.3}) it suffices to restrict the sum in (\ref{5.4}) 
to the lattice sites which are contained in $\R_d-\R_e\,$, i.e. it
suffices to show
\be
\lim_{a\to0}\;\lim_{N\to\infty}\;a^4\sum_{x\in\R_d-\R_e}q^U(x)=I(m_0)Q(A)\,.
\label{5.6}
\ee
Using the locality of $q^U(x)$ in the gauge field (cf. \S3), 
arguments similar to those in \S3 lead to an analogue of (\ref{3.32}):
In the present setting the lattice delta-function has the Fourier 
expansion (analogue of (\ref{3.27}))
\be
\delta_x=\sum_{-\pi\le{}k_{\mu}\le\pi}
\frac{\Delta^4k}{(2\pi)^4}\,e^{-ikx/a}\phi_k
\label{5.6a}
\ee
where $\phi_k(y):=e^{iky/a}$ is the lattice plane wave field with momentum
$k\,$; the domain of the lattice 4-momentum (i.e. the summation domain
in (\ref{5.6a})) in the case of periodic or anti-periodic boundary 
conditions is
\be
\mbox{periodic :}\qquad{}
k_{\mu}&\in&\frac{\pi}{N}\lbrace-N,-N+1,\dots,N-1\rbrace
\label{5.6b} \\
\mbox{anti-periodic :}\qquad{}k_{\mu}&\in&
\frac{\pi}{N}\lbrace-N+1/2,-N+3/2,\dots,N-1/2\rbrace
\label{5.6c}
\ee
and $\Delta^4k=\pi^4/(2N)^4=\mbox{the ``volume per $k$''
in lattice momentum space}$. Using this we find, analogously 
to (\ref{3.32}),
\be
q^U(x)=I(r,m_0,N)q^A(x)+O(a)(x)
\label{5.7}
\ee
where 
\be
I(r,m,N)=\frac{-3r}{8\pi^2}\sum_{-\pi\le{}k_{\mu}\le\pi}
\Delta^4k\,\lambda(k;r,m)\,.
\label{5.8}
\ee
with $\lambda(k;r,m)$ given by (\ref{3.31}).
The $N$-dependence is in the summation domain: (\ref{5.6b}) or (\ref{5.6c}). 
In the $N\to\infty$ limit $\Delta^4k\to{}d^4k$ and we see from
(\ref{5.6b})--(\ref{5.6c}) that in both
the periodic and anti-periodic cases
$\lim_{N\to\infty}\,I(r,m,N)=I(r,m)
=I(m)$ (cf. \S3 for the last equality). It follows that
\be
\lim_{a\to0}\;\lim_{N\to\infty}\;a^4\sum_{x\in\R_d-\R_e}q^U(x)
=I(m_0)\int_{\R_d-\R_e}d^4x\,q^A(x)\,.
\label{5.9}
\ee
By Stokes theorem this reduces to the difference of the integrals
of the Chern-Simons term over the boundaries $\partial\R_d$ and
$\partial\R_e$. The former vanishes since $\partial\R_d$ lies in the 
region where $A_{\mu}(x)$ vanishes. The latter gives the 
the winding number of the map from 
$\partial\R_e\cong{}S^3$ to $SU(n)$ corresponding to the pure gauge 
configuration, and this is precisely $Q(A)$ in the present singular
gauge case. This completes the proof of (\ref{5.3}).

\noindent {\it Comparison with numerical results.} 
Our result (\ref{5.3}) implies that for $0<m_0<2$ the index of $D^U$
reproduces the continuum topological charge $Q(A)$ when $N$ is 
sufficiently large and $a$ is sufficiently small, and that the index reduces 
to $I(m_0)Q(A)$ for general $m_0\not\in\{0,2,4,6,8\}$.
This is something which can be investigated numerically, and in fact
numerical investigations in closely related lattice setups have
already been carried out. 
In finite volume settings such as the present one (where the space of lattice
spinor fields is finite-dimensional) the index of 
$D^U$ equals the spectral flow 
of $-H_m$ as $m$ increases from any negative value to $m_0$ \cite{Neu(PLB)}.
This follows from (\ref{2.11}): $\index\,D^U=-\frac{1}{2}\Tr(\frac{H}{|H|})$
(recall $H=H_{m_0}$) and the fact that the spectrum of $H_m$ is symmetric
and without zero for $m<0$ \cite{ov1,ov2,Edwards}. 
In \cite{Edwards} the spectral flow of $H_m$ was studied numerically
for various smooth $SU(2)$ instanton fields on a finite 
lattice.\footnote{The spectral flow of $H_m$ for various lattice gauge 
fields was first studied numerically in \cite{Itoh}.}
One of the situations considered in \cite{Edwards} 
was instanton field in a singular
gauge with anti-periodic boundary conditions on $H_m$. I.e. anti-periodic
boundary conditions on the lattice spinor fields, and the lattice transcript
of the instanton field modified at the boundary links to make it 
anti-periodic, so that $H_m$ is a well-defined hermitian operator on these
spinor fields. This is a minor modification since the localisation radius
of the singular gauge instanton 
in \cite{Edwards} was well within the region covered by the lattice, i.e. it
was almost vanishing at the boundary of the lattice.
This is similar to the setup that we have considered in this section:
the singular gauge instanton can be approximated by a gauge field
$A_{\mu}(x)$ with singularity which is pure gauge in a neighbourhood of
the singularity and vanishes outside a bounded region.
The results of \cite{Edwards} are compatible with ours:
They numerically determined the crossings of the origin by low-lying
eigenvalues $\lambda(m)$ of $H_m$ for $0<m<2$ and found precisely one 
crossing, which occurred reasonably close to zero
(at $m\approx0.5$), and the slope of $\lambda(m)$ at
the crossing was such that the spectral flow of $-H_m$ from this crossing
agreed with the topological charge of the 
instanton (cf. Fig. 7 of \cite{Edwards}). This can be equivalently expressed
as $\index\,D^U=Q(A)$ for $0.6<m_0<2$. This is an approximative
numerical confirmation of our analytic result in this section.
Numerical studies of the spectral flow of $H_m$ in other situations 
in \cite{Itoh,Edwards}, or equivalently, numerical studies of $\index\,D^U$ 
as a function of $m_0$ \cite{Chiu(PRD1+2)}, are also compatible with the 
the classical continuum limit result for $\index\,D^U$ that we have
shown for the particular situation considered in this section (and which
has also been analytically shown in the finite volume torus case
in \cite{DA(torus)}): For ``sufficiently smooth'' lattice gauge fields
it was found that (i) crossings of the origin by eigenvalues $\lambda(m)$
of $-H_m$ occur close to $m=0,2,4,6,8\,$; if the net spectral 
flow from crossings close to $m=0$ is $Q$ then the net spectral flows
from crossings close to $m=2,4,6,8$ are, respectively, $-4Q\,$,
$6Q\,$, $-4Q\,$, $Q\,$, 
and (ii) if $U$ is the lattice transcript of a continuum
field $A$ then $Q=Q(A)$. (A situation where this does {\it not} hold
is when $U$ is the lattice transcript of an instanton field in a
regular gauge and anti-periodic boundary conditions are imposed on
$H_m$ \cite{Edwards}. This is to be expected though, since in this case
the instanton field is approximately pure gauge at the boundary of the region
covered by the lattice, and is therefore not close to satisfying 
anti-periodic boundary conditions.)

\section{Conclusion}

In this paper we have rigorously verified that the index density/axial anomaly
for the overlap Dirac operator has the correct classical continuum limit
in the infinite volume lattice setting, and in a finite volume setting
where the the continuum limit involves an infinite volume limit.
The only condition on the continuum field $A_{\mu}(x)$ required to establish
this result is that the plaquette variable of its lattice transcript
satisfies 
\be
||1-U(p)||<\e(m_0)
\label{7.1}
\ee
for sufficiently small lattice spacing $a$, with $\e(m_0)$ small enough 
to guarantee the existence of a lower bound $b(m_0)>0$ on $H^2$
\cite{local,Neu(bound),DA(bound)}, which in turn guarantees the locality
of $D^U$ in the gauge field \cite{local}. The locality plays a central
role in our arguments. The condition (\ref{7.1}) is automatically
satisfied for small $a$ when $A_{\mu}(x)$ and its first and second order 
partial derivatives are bounded on ${\bf R}^4$, since it can then be shown 
that $||1-U(p)||\sim{}O(a^2)$. However, using the locality property
we have seen that the condition is 
satisfied for more general gauge fields, including some cases where the field
has a singularity or diverges at infinity. In particular our results cover
the case of topologically non-trivial gauge fields on ${\bf R}^4$.
This case was was not covered by previous works on this topic (cf. the 
discussion in \S2).
Our approach was to decompose the operator $H^2$
appearing in the overlap Dirac operator 
$D=\frac{1}{a}\Bigl(1+\g5\frac{H}{\sqrt{H^2}}\Bigr)$ into $H^2=L-V$
where all the gamma-matrices are contained in $V$ and $||V||\sim{}O(a^2)$.
Then, starting with an integral representation for the inverse square root, 
we expand $1/\sqrt{H^2}$ in powers of $V$. It was found that only the term
of order 2 in $V$ contributes to the index density 
$q^U(x)=-\frac{1}{2}\tr\frac{H}{\sqrt{H^2}}(x,x)$ in the classical continuum
limit, and the contribution from this term was shown to reduce to the
continuum density $q^A(x)$ (for $0<m_0<2$).

The expansion of the overlap Dirac operator in powers of $V$ is of interest
in its own right and may have other useful applications. Unlike an expansion
in powers of the gauge field, such as the one given in \cite{KY}, which
requires the gauge field to be small (i.e. link variables close to 1),
our expansion only requires a bound of the form (\ref{7.1}) to hold,
i.e. plaquette variables close to 1, and can therefore be used for 
topologically non-trivial gauge configurations.

Subtleties/pitfalls which arise when considering the index of the overlap
Dirac operator in the infinite volume setting were discussed (\S4).
In this case the index is generally ill-defined a priori. Formally we
have $\index\,D^U=a^4\sum_xq^U(x)$, and we pointed out that the previous 
result $q^U(x)\to{}q^A(x)$ for the index density does {\it not} imply
$a^4\sum_xq^U(x)\to\int_{{\bf R}^4}d^4\,q^A(x)=Q(A)$ in general (as was
erroneously claimed in \cite{K}):
The argument was seen to break down due to the infinite sum over lattice
sites in $\sum_xq^U(x)$.
We showed however that $a^4\sum_xq^U(x)\to\int_{{\bf R}^4}d^4x\,q^A(x)=Q(A)$
in the case where $A_{\mu}(x)$ is pure gauge outside a bounded region
of ${\bf R}^4$. This is a first step towards showing the following
conjecture: $\index\,D^U$ is well-defined and equal to $a^4\sum_xq^U(x)$
when the lattice field $U_{\mu}(x)$ converges sufficiently quickly
to a pure gauge configuration at infinity, and in this case 
$\index\,D^U\to{}Q(A)$ in the classical continuum limit. This conjecture
(the continuum analogue of which is known to be true \cite{Uhlenbeck})
remains as an interesting problem for future work.

In order to verify that the index of the overlap Dirac operator really is
able to capture topological information we considered (in \S5) a finite
volume lattice setting where $\index\,D^U$ is well-defined from the beginning,
namely when $U$ is the lattice transcript of a continuum field $A_{\mu}(x)$
which is in a singular gauge and vanishes outside a bounded region.
(For technical reasons we also required $A_{\mu}(x)$ to be pure gauge
in a neighbourhood of the singularity. We described how such gauge fields
can be readily constructed from maps $\phi:S^3\to{}SU(n)$.)
We showed that $\lim_{a\to0}\,\lim_{N\to\infty}\,\index\,D^U=Q(A)\,$, 
where $N\to\infty$ is the infinite volume limit. This result complements
a previous numerical result obtained in a similar setting in \cite{Edwards}.

For the finite volume 4-torus case the results 
$\lim_{a\to0}\,q^U(x)=q^A(x)$ and $\lim_{a\to0}\,\index\,D^U$ $=Q(A)$
are established for general gauge fields in a separate paper
\cite{DA(torus)}. In that case the result for $q^U(x)$ immediately implies
the result for $\index\,D^U$ since the volume is finite. However, the finite
volume causes additional techinical complications in the calculation of 
$\lim_{a\to0}q^U(x)\,$; we refer to \cite{DA(torus)} for the details.

Finally we mention that, while the settings we have considered were all
in 4 dimensions, everything generalises straightforwardly 
to arbitrary even dimension
$2m$, with gauge group $SU(n)$ for $2m\ge4$ and gauge group $U(1)$ 
in the 2-dimensional case. In the latter case our result $q^U(x)\to{}q^A(x)$
confirms a numerical result in \S10 of \cite{ov2} where the axial anomaly
$\A^U(x)=2iq^U(x)$ was numerically determined and compared to the
continuum axial anomaly for a particular topologically non-trivial
$U(1)$ gauge field on the 2-torus.\footnote{I thank Herbert Neuberger
for pointing this out (cf. \cite{Neu(Taiwan)}).}

{\it Acknowledgements.} I am grateful to Martin L\"uscher for his 
feedback on a previous version of this paper, and for his suggestions.
I also thank Ting-Wai Chiu and Herbert Neuberger for 
discussions/correspondence at various stages. The author is supported by 
an ARC postdoctoral fellowship.

\pagebreak

\end{document}